# Quorum sensing and remote synchronization in networks of Kuramoto oscillators: a biological interpretation

Vincenzo Fioriti

*Deep Blue Srl, Piazza Buenos Aires, 00198 Rome, Italy*
*email to: vincenzo.fioriti@dblue.it*

*Abstract* - Non-linear oscillator networks have revealed properties as the remote synchronization and the quorum sensing. The remote synchronization, defined as the synchronization of nodes not directly connected by any sequence of synchronized nodes, was found firstly in networks of amplitude oscillators and recently in bipartite delayed networks of phase oscillators. The quorum sensing, a biological information scheme discovered in cell aggregates, has been investigated in amplitude oscillators coupled by a common medium. Implications of such findings are important in technology and biology. We show both of them in non-bipartite, biologically plausible networks of Kuramoto oscillators. The quorum sensing emerges using the graph edge density, while the remote synchronization is obtained by means of an oscillator acting as a pacemaker. In the remote synchronization two distinct groups of well inter and infra-synchronized nodes, separated by non-synchronized paths, appear clearly. Our biological interpretation is that the remote synchronization, bypassing the normal quorum sensing mechanism, is responsible of the pathological cell proliferation. This approach seems suitable to study the quorum sensing alterations due to genetic mutation or to the environmental action before the actual mass replication begins.

## I. INTRODUCTION

One of the simplest forms of information elaboration in Nature is the periodic variation of a quantity. Chemical, biological, physiological, physical oscillators are everywhere, regulating the everyday life with impressive precision. When coupled each other to set up a network, the collective dynamics reveals chaos, synchronization, randomness, periodicity, sometimes in close connection in the same experimental or simulative environment. Usually the analysis is interested in the directly linked oscillators, taking for granted that the synchronization action is to be transferred by the links, i.e. a local synchronization is governing the whole process of synchrony. However, during the last years the *remote synchronization* (RS) has been discovered in Stuart-Landau amplitude oscillators networks [1]. RS is defined intuitively as the synchronization between nodes that are not connected by any path of synchronized nodes. The synchronization is measured by the Kuramoto order parameter $R(t)$ ranging $0 < R(t) \leq 1$ [2], with a threshold set to $R(t) = 0.8$ [1]. Recently, [3] have shown a similar phenomenon in bipartite phase oscillator graphs with time delays, thus the RS is more general then suspected, but still limited to a particular type of graph. In this work we extend the RS to non-bipartite phase oscillator graphs with a focus on the biological plausibility required to study the basic properties of the *quorum sensing* (QS). The quorum sensing is a biological distributed computing scheme, observed in populations of bacteria and cells. Cells release diffusible signal molecules (the so-called autoinducers), which by accumulating in the environment induce population-wide changes in gene. The bio-chemical details are complex, hence we reduce roughly the mechanism to the dynamical action of a protein receptor of the autoinducers (AI) and to the autoinducers [4, 5]. Cells try to synchronize their actions despite random disturbances, here represented as phase frustrations, noise and the physiological removal of the AI. During the synchrony periods, receptors can bind the AI more easily, triggering a further production of AI and so on, in a positive feedback loop.

This ability of diffusing QS signals to synchronize gene expression in spatially extended colonies is not completely clear [6, 7], nevertheless some efforts in modelling numerically the QS through oscillators coupled by a common medium have been made with success [8, 9], at least for the very basic characteristics. The governing parameter is the oscillator population density, producing a density dependent synchronization threshold similar to a first order state transition. Note that in [8, 9] only chaotic or amplitude oscillators have been used, instead the QS in Kuramoto phase oscillator networks (KM) was not addressed directly, despite it is the most investigated model of non-linear oscillator. However, in the standard KM exists a first order transition to synchronization called explosive synchronization (ES), correlated to the node degree [10, 11, 12, 13, 14, 16]. This correlation suggests that, at least roughly, a QS-like effect could be present also in Kuramoto oscillator graphs where the governing parameter is the graph edge density.

The main tool is the heterogeneous Kuramoto model:

$$d\theta_i/dt = \omega_i + K_i/N \sum_i a_{ij} \sin(\theta_j - \theta_i) \qquad i, j = 1, ..., N \qquad (1)$$

where $\theta_i(t)$ are the phase variables, $\omega_i$ the intrinsic or natural frequencies, $K_i$ the couplings, $a_{ij}$ is 1 or 0 depending on the existence of an edge between node $i$ and node $j$ (therefore we have local connections, as opposed to the all-to-all connections), $N$ is the number of oscillators or nodes. Above the critical value of the couplings $K_c$ the oscillators phase synchronize:

$$\lim_{t \to \infty} (\theta_j - \theta_i) = 0$$

The synchronization is measured by the standard order parameter, defined as:

$$R(t) = 1/N \left| \sum_j e^{i\theta_j(t)} \right| \qquad \text{with} \quad i = \sqrt{-1}, \quad j = 1, ..., N \qquad (2)$$

Authors of [11, 12, 13] find the explosive synchronization in presence of the frequency assortativity (a proportionality relation between the natural frequency $\omega_i$ and the node degree $d_i$) or in presence of a large average degree such as in Barabasi-Albert graphs. In the first case, we have a simple proportionality relation, where $c$ is a constant

$$\omega_i \sim d_i/c \qquad (3)$$

and *i* may cover only a subset of the nodes. In other words to have assortativity it suffices only a small number of nodes [11], indicating that the ES may be triggered by a *local* increase of the edge density and does not require a global property.

In the second case, the critical coupling strength is inversely dependent from the average degree of the network $<d>$:

$$K_c = 2/\pi <d> g(<d>) \qquad (4)$$

and above the critical value $K_c$, the synchronization explodes.

A number of interesting papers have been written on the issue. In [12] the frequency-degree assortativity induces chaos in the Kuramoto order parameter fluctuations, both in large 1000 nodes networks and in the reduced 20 nodes networks via the Ott-Antonsen ansatz [15]. Therefore, an assortative Kuramoto model is able to show chaotic variations in $R(t)$, just like in the amplitude oscillators population density case [8], but gaining the advantage of saving cumbersome calculations when $N \to \infty$. A second, more important advantage, is that the KM represents the simplest non linear oscillator equipped with an extensive mathematical treatment. Anyway, it should be remembered that at the weak coupling limit, the Stuart-Landau model reduces to the Kuramoto model. [14, 19] extend the first order transition to various distributions of $\omega_i$ correlated to the degrees as a *generic property* of phase oscillator network. In the same paper it is shown the effect of an external pacemaker with amplitude $A_0$

$$d\theta_i/dt = \omega_i + K/N \sum_i a_{ij} \sin(\theta_j - \theta_i) + A_0 \sin(\theta_p - \theta_i) \qquad i, j = 1, ..., N \qquad (5)$$

on the synchronization, resulting in a so-called magnetic state after the external action is removed, while in [17] the pacemaker is shown to influence heavily the synchronization rate. The internal pacemaker is assessed by the Authors of [18], who suggest that a gradient in the natural frequency distribution generates spontaneously a pacemaker effect. Again, it is interesting that only a subset of oscillators is needed to produce the self-organized pacemaker.

In this paper we show firstly how the variation of the global edge density $\rho$ in Erdos-Renyi graphs mimics the QS. We use the term "graph" in the mathematical sense and "network" as the physical implementation of the graph; basically the meaning is the same. Then we will assess the remote synchronization in a small non bipartite graph between two node subsets (called driver-response), connected by a non-synchronized bridge-node acting as a pacemaker. Finally, a biological interpretation of the results is given.

## II. THE SIMULATION FRAMEWORK

In our simulation framework, $N$ is the number of cells, the global edge density is the AI density, the degree represents the local AI density, the phase variables are the receptor proteins, the natural frequencies are the autoinducers production rates [4, 5], frustrations simulate the natural noise and delays, the global coupling $k$ represents other bio-chemical parameters, the forcing is the environmental action. We understand that two cells (nodes) are linked by an edge if an AI from a cell activates the receptor of the other. It was suggested [20, 21] that the spatial distribution of cells is as important for sensing at least as the cell population density, therefore we use the edge density to take into account the cell-to-cell spatial relations.

**A. Edge density and synchronization**

We use the Erdos-Renyi undirected pseudo-random graphs (ER) to study the influence of the edge density $\rho$ on the synchrony. The ER model starts with $N$ nodes and connects each pair of nodes with a certain probability. The node degrees follow a Poisson distribution, which indicates that most nodes have approximately the same number of links close to the average, while the nodes deviating from the average are rare [16].

When $\rho$ increases, degrees get more and more similar, and finally for $\rho \approx 1$ all nodes have almost the same degree. If the frequency assortativity holds [14], also the natural frequencies $\omega_i = d_i/c$ will tend to a unique value, fostering sufficiently the synchronization even below the critical coupling.

In ER graphs the initial values of the degrees is low (low $\rho$), hence no synchronization is possible when $K < K_c$, but when the degrees are almost equal (high $\rho$), then we have a synchrony state. In the biological interpretation the abundance of AI (that we represent as a high edge density) getting close to $\rho = 0.9$ triggers a QS-like synchronization (Fig. 1). Note that a uniform distribution of the natural frequencies $\omega_i$ or an excessively large frequency value would not achieve this QS-like behaviour.

The mathematical model is the Kuramoto equation (1) with the order parameter (2), natural frequencies with frequency-degree proportionality $\omega_i = d_i/c + \varepsilon(<d>)$ (frequency assortativity with a small gaussian noise depending on the average degree), one weak coupling $K = 0.02$, $c = 70$, initial conditions on $\theta_i(0)$ uniformly distributed inside the interval $[-\pi, \pi]$, no frustrations or pacemaker. The networks tested are Erdos-Renyi pseudo-random graphs with $N = 21$ and $N = 63$ nodes.

After various runs and 600 data transitory, $<R(t)>$ time averages are calculated. In Fig. 1a an abrupt transition to synchrony is evident beyond $\rho = 0.9$ and so in Fig. 1b. The QS-like behaviour is sensitive to noise in large sized networks, preventing a full synchronization and causing oscillations to the average order parameter (see Fig. 1b), nevertheless, the QS phenomenon has been replicated.

ER graphs may not be the most correct model of the spatial links, since representing spatially cell aggregates by means of graphs is unusual and furthermore indications in the literature are just a few. In fact, the reductionism, the prevalent mainstream research trend in the last decades, has focused on individual molecules rather than their complex interaction, that is better described by graphs [16]. On the other hand, a discussion on this issue is out of the scope of the paper and anyway the ER graph seems a reasonable approximation of reality, if we want to test only the synchronization characteristics.

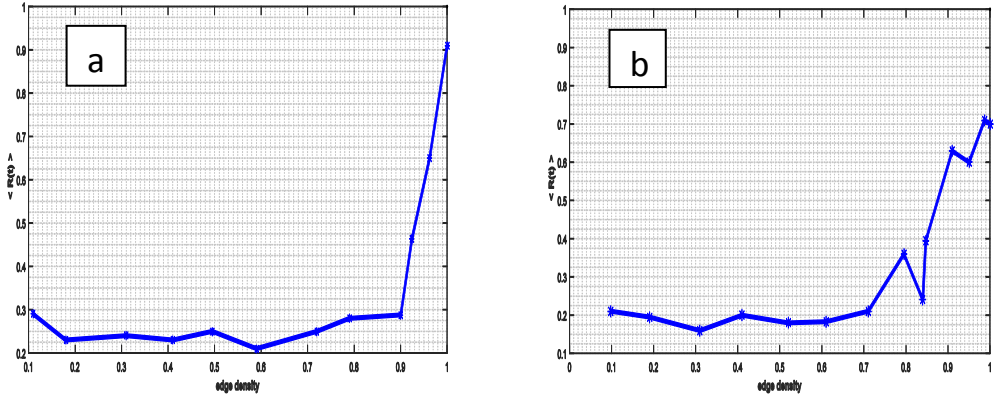

**Figure 1.** The global edge density $\rho$ vs. the averaged order parameter $<R(t)>$ for Erdos-Renyi pseudo-random graphs of 21 and 63 nodes. a) an abrupt transition to synchrony is apparent beyond $\rho = 0.9$ in the 21 nodes graph. b) the graph of 63 nodes produces a transition beyond $\rho = 0.85$, but noise depending on the average density reduces the order parameter synchronization value to 0.7.

### B. The remote synchronization

To draft a more biologically plausible framework without getting involved in a detailed modelling of bio-chemical aspects, we have selected some features for the KM to be implemented in the toy network of Fig. 2. For example, frustrations are necessary because random phase drifts have been actually observed in natural aggregates [4]. We add noise to the natural frequencies and frustration to phases to be more close to reality, observing that the frustrations $\varphi_{ij}$ translate the time delays into corresponding phase offsets [22]. In a plausible model one expects different natural frequencies, noises, frustrations, delays, different weak couplings, local connections and the capability to show dynamic behaviours such as periodicity, chaos, randomness. Also the influence of the environment has to be considered: the simplest way is to use a forcing/pacemaker action entrained with the oscillators representing the cells. Our network of Kuramoto oscillators in Fig. 2 is a simple, undirected, non-bipartite graph of 21 nodes, intuitively grouped into three subsets (driver, channel, response). Of course, this is only a toy graph to demonstrate the RS more easily than using a large graph with a quasi-random topology.

The Kuramoto model with a pacemaker is:

$$d\theta_i/dt = \omega_i + (K_i/N) \cdot \Sigma_i a_{ij} \cdot \sin(\theta_j - \theta_i - \varphi_{ij}) + A_0 \cdot \sin(\Omega_0 t - \theta_i) \qquad i,j = 1, ..., N \qquad (6)$$

$$\omega_i = d_i/c \quad \text{and} \quad K_i = k \cdot d_i \qquad (7)$$

where $K_i$ are the local couplings, $k = 0.02$ is the global coupling (well below the critical coupling), $d_i$ the degrees, frustrations $\varphi_{ij}$ are uniformly distributed the interval $(30/100)\cdot[0.15, \pi\cdot 0.135]$, initial conditions $\theta_i(0)$ are uniformly distributed in $[-\pi, \pi]$, the constant $c$ varies in $[11, 20]$ (we choose $c = 15$), $A_0 = 0.01$, the Runge-Kutta simulation step is 0.4.

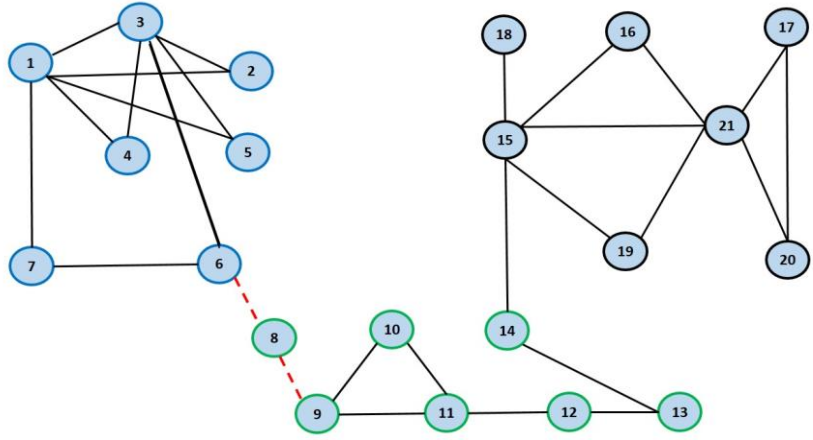

**Figure 2. The network of Kuramoto oscillators. The blue circled nodes are the driver set (1-7), green the channel (8-14), black the response set (15-21). Node 8 with dotted edges is the bridge that isolates the two groups. The edge density of this non bipartite graph is less than 0.15.**

The pacemaker depends on the node 8, although other choices are possible as well. Adding the noise $\xi$ to the frequencies:

$$\Omega_i = \omega_i + \xi_i \tag{8}$$

with $\xi_i$ uniformly distributed in the interval $\xi \in [\,[\,0.01,\,0.05]$ , (9)

if $\Omega_0 t \approx \theta_8 + \beta$ , where $\beta \in [\pi/10\,,\,\pi]$ is a perturbation, then (6) becomes: (10)

$$d\theta_i/dt = \Omega_i + (K_i/N)\cdot\Sigma_i\, a_{ij}\cdot\sin(\theta_j - \theta_i - \varphi_{ij}) + A_0\cdot\sin(\theta_8 - \theta_i + \beta) \tag{11}$$

Node 8 was selected since is the bridge between the driver-response groups, but other choices among the channel nodes are possible; it suffices that the pacemaker node $h$ of the type $-A_0 \cdot\sin(\theta_h - \theta_i)$ belongs to the channel: $h \in \{8,9,10,11,12,13,14\}$. Differently from [18] the pacemaker node has not the highest natural frequency, rather it is lower than the average natural frequency.

In other words, when an oscillator accidentally begins to act as a pacemaker or an external forcing is entrained with an oscillator comes into play the model (8-11), generating the remote synchronization. The pacemaker action produces a sort of *negative* feedback on all the nodes allowing the synchronization of the driver-response groups, but preventing the oscillator $h$ (in this case $h = 8$) to synchronize with its direct neighbours. Note that using the external pacemaker $\pm A_0 \cdot\sin(\Omega_0 t)$ or a positive feedback, no remote synchronization would result.

It was demonstrated in experiments with hundreds of synthetic genetic oscillators that if the period of the forcing is sufficiently close to the natural period of an oscillator, the oscillator can be entrained [23], with an almost fixed phase difference. In synthetic genetic oscillating aggregates, the entrainment may produce the Arnold tongues [23] while the global intercellular coupling give rise to synchronized oscillations [24].

Frustrations cause the interaction functions to pull the phases away from complete synchrony, even when the natural frequencies are identical [22], but at the same time force directly connected oscillators to maintain a constant phase difference, reinforcing the overall synchronization [25], and, most of all, makes more biologically plausible the model (8-11).

Results over several runs show a very high synchronization between the driver-response groups while the node 8 is completely desynchronized with its direct neighbours, producing a synchrony gap between the two groups. Therefore a remote synchronization is operating. Moreover, $R_d(t)$ and $R_r(t)$ are linearly correlated with a large correlation coefficient, see Fig. 3.

In the following, the simulation parameters are as above, moreover $\beta = \pi$, $c = 14$ (different values of $\beta \in [\pi/10, \pi]$ and $c \in [11, 20]$ do not alter the results). After the transitory, the results of the numerical simulation are:

$<R(t)> = 0.792$ is the average overall order parameter $R(t)$ (see Fig. 5),

$<R_d(t)> = 0.877$, $<R_m(t)> = 0.688$, $<R_r(t)> = 0.830$,

are respectively the channel, the driver, the average response order parameters, (Fig. 3). The overall synchronization is lower with respect to the response and driver individually considered. The infra-synchronization of the driver and of the response is high, even better than the whole network.

$<R_{d\,r}(t)> = 0.853$, $corr(R_d, R_r) = 0.955$,

Above the driver-response order parameter, (Fig. 4) and their correlation coefficient. The node 8 is clearly not synchronized with its neighbours 6 and 9:

$<R_{6,8}(t)> = 0.174$, $<R_{8,9}(t)> = 0.090$,

producing a gap between the driver group and the response group (Fig. 1), since the path is unique.

It is well known that the limited size of the graphs generate fluctuations in the order parameter. Due to the fluctuations, for weak couplings the incoherent state order parameter is close but not exactly zero, while during the full synchronization is not exactly 1. Here we obtain the thresholds $R_{dr}(t) \approx 0.8$ and $R_{dr}(t) \leq 0.4$ of the coherent/incoherent state, more conservative with respect to the RS definition given in [1].

When the intrinsic pacemaker in (11) is turned off, synchronizations and correlations disappear even increasing the overall coupling $k$ ten times; moreover, for $\beta > \pi$ the overall synchronization remains, but the remote synchronization deteriorates.

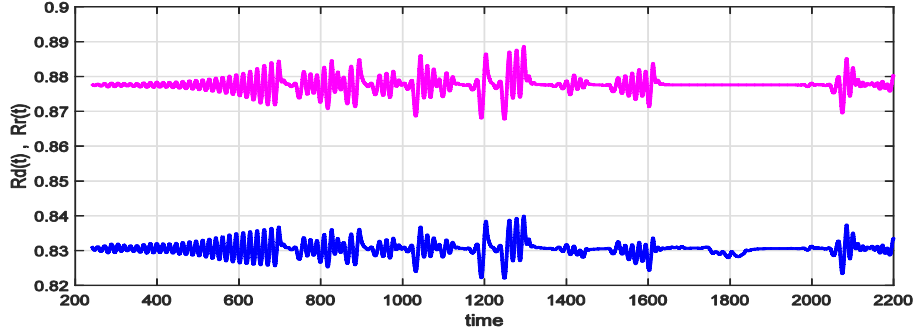

**Figure 3.** Order parameter $R_d(t)$ of the driver (blue, bottom) and $R_r(t)$ of the response group (red, on top) considered individually for $\beta = \pi$. Their correlation coefficient is $corr(R_d, R_r) = 0.955$.

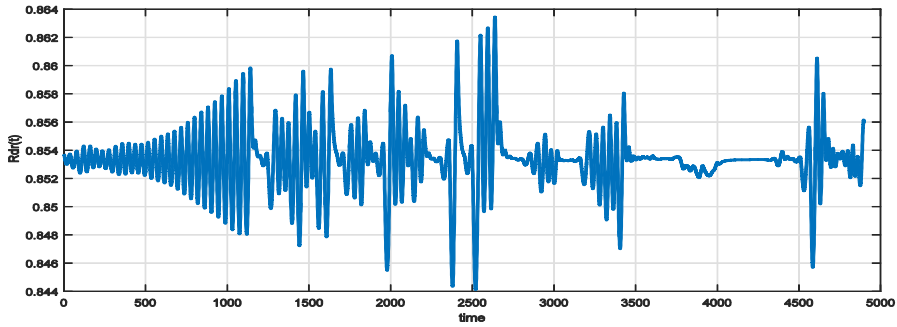

**Figure 4.** Order parameter $R_{d\,r}(t)$ of the driver-response group as a whole (time resolution doubled).

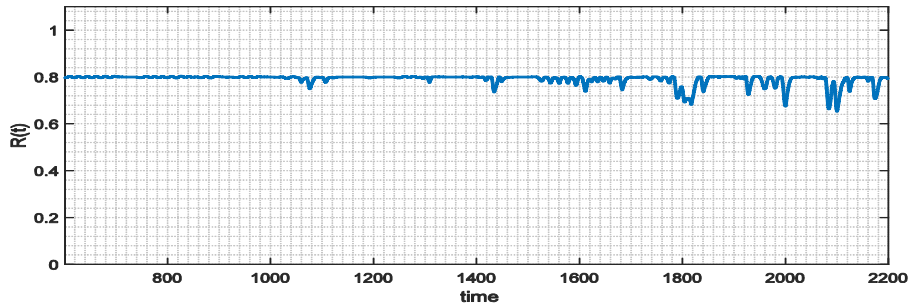

**Figure 5.** The variation of the order parameter $R(t)$ of the whole graph of Fig. 2.

Interestingly, the assortativity relations (7, 8) are not necessary for all the nodes (although the RS *quality* depends on the full assortativity). In fact, results hold also if (7) is applied only to the driver group, confirming the analogous conclusions of [11, 18, 19]. For example, [11] restrict the degree-frequency correlation to the hubs promoting the synchronization in a typical biological benchmark network, the neural network of the worm *Caenorhabditis elegans*.

Finally an observation: the Kuramoto model was not reputed able to activate the RS because of the low strength of the amplitude [1]. However the Kuramoto order parameter can be seen as a kind of AM modulation [26], resulting from the interaction of an external periodic signal with local phase oscillators. The periodic signal offers the energy, the local oscillators the information to be transmitted.

**C. Discussion**

The standard Kuramoto model was modified in order to resemble, roughly, a model of a biological environment to achieve the synchronization through the quorum sensing mechanism based on the edge density and the remote synchronization via pacemaker. It was shown in Fig.1 a QS-like mechanism in small ER graphs, governed by the edge density instead of the population of oscillators. The remote synchronization is based mainly on the pacemaker action; the remote and separated groups of oscillators may reach high levels of coherence if one oscillator is able to work as an intrinsic pacemaker *or* the external environment provides a periodic stimulus. The remote synchronization represents therefore a communication channel alternative to QS, activated at a low density regime by a global feedback. Hence, on the one hand the QS cannot push towards the synchronization because of the low edge density and of the presence of the frustrations, but on the other hand the RS is capable to force the synchronization, bypassing the QS via the pacemaker.

This "alternative" QS could be useful to model the metastasis as in pathogenic bacteria to determine when their population number is high enough to collectively form biofilms in host organisms [27, 28, 29]. The RS could be considered a *pathologic* QS reducing the local communications, but enhancing the remote one, without relying on the AI density. In this regard, the case of the oscillator acting as a pacemaker resembles the genetic mutation of a single cell, while the external periodic stimulus represents the influence of the environment on the aggregate.

The genetic mutation of an oscillator concerns also its topology, i.e. the connections of the mutate oscillator in the sense that every oscillator is linked to the mutate oscillator. This situation could be caused by an abnormal production of AI from the mutate oscillator; another mechanism able to generate an intrinsic pacemaker is a gradient in the distribution of the natural frequencies [18]. In both cases a global pacemaker signal is obtained, hence this interpretation combines the major known causes of carcinogenesis, i.e. mutation and environment.

It is important to observe that the model described above pertains the situation *before* the cell proliferation, since the number of *nodes* in the network does not vary. However, when the synchrony is well established, any variation of the oscillators' number in one of the two synchronized groups has to be followed by an analogous variation of the other, because of the linear correlation of $R_d(t)$ and $R_r(t)$, according to (2). Therefore, if for any reason the number of oscillators in the driver increases, also the number of oscillators of the response increases and vice-versa; at this point, the mass replication is turned on.

## III. CONCLUSIONS

We have shown the existence of the remote synchronization in generic non bipartite graphs via pacemaker and of the quorum sensing in Erdos-Renyi graphs. Previously these two phenomena were not discussed explicitly in the literature of Kuramoto oscillators, despite a very large number of papers about the non linear dynamics in complex networks. We have focused on the strict coherence of the remotely synchronized oscillators, manifested in the high linear correlation between the driver-response order parameters, mimicking an amplitude modulation between the two. Moreover, we have found the QS in Erdos-Renyi phase oscillator graphs using the edge density as control parameter (although differences with the performances of the Landau-Stuart amplitude oscillators persist). Hence, the Kuramoto model seems a good candidate to model biological phenomena, in particular the cell proliferation before the actual mass replication begins. In this sense, we understand the remote synchronization among nodes as a pathologic form of quorum sensing, meaning that the replication signal expressed through the common synchrony occurs without the previous growth in cell density required by the physiological QS. Once the synchrony is established, a possible small local cell increase is replicated remotely, because the synchronization forces the direct proportionality of the populations and successively a positive loop might trigger the mass increment of both populations.